# Universality classes of surface wave turbulence as probed by laser Doppler velocimetry in viscous fluids.

Mikheil Kharbedia[1] and Francisco Monroy[1,2,*]

[1]*Department of Physical Chemistry, Universidad Complutense de Madrid,
Ciudad Universitaria s/n E28040 Madrid (Spain).*
[2]*Unit of Translational Biophysics, Instituto de Investigación Sanitaria Hospital
Doce de Octubre, E28041 Madrid (Spain).*

By taking advantage of laser Doppler velocimetry (LDV), we have explored the existence of discrete wave cascades on fluid interfaces excited upon monochromatic excitation. We have studied a set of viscous liquids of variable capillarity spanning a broad range of frictional stresses. The existence of an inertial regime of weak surface wave turbulence (weak-SWT) has been pointed out as cascades of discrete harmonics with amplitudes compatible with universal Kolmogorov's scaling. The transit from weak- to strong- surface turbulence has been also explored revealing the importance of inertia over surface forces to describe the crossover. LDV has been proven as a powerful tool to probe surface waving motion.

## I. INTRODUCTION

Fluid turbulence is a nonlinear hydrodynamic state that sustains far-from equilibrium flows under the action of an external driving force [1,2]. A prominent feature of turbulent flows is the appearance of a cascade of inertial (frictionless) motions produced under coupling between the macroscale driving source and a sequence of non-linearly excited modes at smaller scales [2-5]. A hierarchy of scales appears in this non-linear flow, so turbulence emerges as an ensemble of many-scale motions, which constitute an inviscid cascade —the so-called inertial interval [2]. At scales well within this inertial domain, the properties of the cascade are expected to be universal [6]. Finally, at a characteristic failure scale, where the viscous stresses are the sharpest to overcome the inertial response, the cascade undergoes a "ultraviolet catastrophe" [1,2]. The physical simplicity inherent to the inertial interval was early envisaged by Kolmogorov as a self-similar scaling [2], which can be described in terms of scale invariance and universality [3-7]. Today, the scaling universality inherent to the inertial interval has ransomed the Kolmogorov's notion of synergistically organized cascades as the structural "skeleton" of turbulent flows [6,8]. Furthermore, the breakdown of Kolmogorov's symmetries has inspired the notion of criticality as a pathway from universal turbulence to chaotic flow [6,9]. Building upon variational principles similar to those governing thermal equilibrium [10], broken symmetries are nowadays envisaged as an opportunity to identify the scaling laws that govern the statistical mechanics of far-from-equilibrium flows [6,9]. The theories of universal weak turbulence (weak-WT) have been fruitfully exploited to study many natural phenomena, such as ocean waves [11], atmospheric streams [12] and interstellar plasmas [13,14], among others. In the laboratory setting, weak-WT is used to recreate some turbulence processes relevant in technology [15,16], and to detect possible anomalies from predicted universal behavior [17-19].

Wave turbulence (WT) can be excited as a set of nonlinear waves giving rise to turbulent flow far away from thermal equilibrium [4,5]. The phenomenon of weak-WT is recognized as a perturbative hydrodynamic status that pumps interacting nonlinear waves by frictionless resonance with the forcing mode [3-5]. Specifically, the universality class of weak-WT is amply known as an inviscid ensemble of collective excitations (see Fig. 1A), which are identified as wave resonances satisfying energy conservation [20]. WT-cascades can be characterized by their power spectra $P(\omega)$, which represent the amplitudes of the different modes of frequency $\omega$. Building upon the Hamiltonian formalism for waves in continuous media, Kolmogorov's theory describes the weak WT-cascades with a spectral density decreasing with frequency as a power law $P(\omega) \sim \omega^{-\alpha}$, with scaling exponent ($\alpha$) representing the universality class of the waving ensemble as defined by the symmetry of the Hamiltonian [2-4].

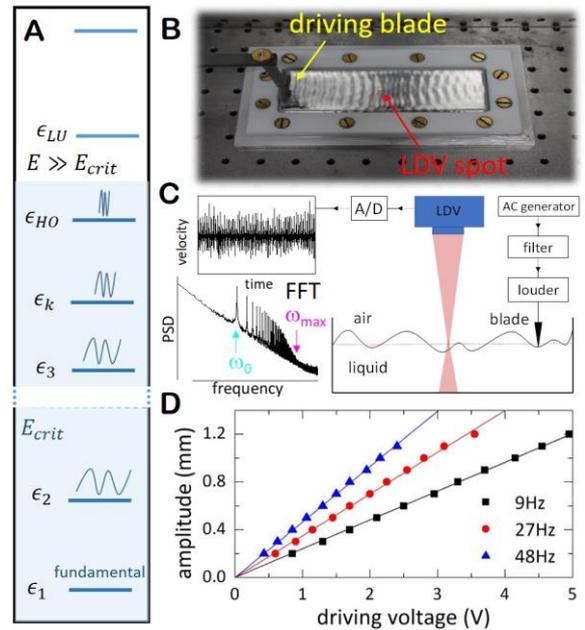

**FIG. 1. A) WT ensemble** representing the interacting nonlinear waves as interacting microstates that constitute the inertial cascade of weak turbulence (see Section II for a discussion). **B) SWT experimental set-up** exciting the ensemble of nonlinear planar waves from a monochromatic forcing deformation exerted in the surface by a vertical blade. A Doppler velocimeter impinging a laser beam in the normal direction is used to measure the vertical velocity of the surface at a point distant from the exciting line (see the red spot). **C) Schematic of the device** used to excite/record SWT cascades at air/liquid interfaces. The forcing blade is powered from a AC synthesizer, whose monochromatic exciting signal is further filtered through a high-gain amplifier. The analogic signal recorded from the LDV-meter is A/D-converted as a time-trace, which become a treatable dataset for FFT from which the spectral profile of the discrete cascade is obtained (see the discrete modes starting at $\omega_0$ and terminating at $\omega_{max}$). **D) Linearity** of the exciting device with the RMSD-voltage of the AC-driving signal at different excitation frequencies in the working range.

Surface wave turbulence (SWT) occurred at fluid interfaces is a genuine surface response to oscillatory forcing against the restoring forces imposed by surface tension and gravity [21]. The adequate analytic framework is provided by the Kolmogorov-Zakharov (KZ) theory [4,5], which is formally grounded on the scaling theory of weak turbulence [2,3]. For monochromatic excitation above the coupling threshold $E_{crit}$, KZ theory predicts a cascade of discrete surface modes describable by Kolmogorov's scaling laws [4,5,21]. The existence of collective excitations on the free surface of different liquids has been experimentally verified [22-28], even with quantum fluids [29]. However, most of those studies were performed under broadband and/or random excitation giving rise to continuous cascades [22-25]; only a couple deal with monochromatic excitation [26,27]. Nowadays, a gap exists indeed on the phenomenological understanding of discrete SWT cascades, which represent the adequate setting for tests of universality. Accurate SWT-probing in laboratory experiments requires both, excitation of surface waves under controlled external forcing and *in situ* measurements of intrinsic flow velocities, a goal only achievable by velocimetry techniques such as particle image velocimetry (PIV) [30], and laser Doppler velocimetry (LDV) [31]. Particularly, LDV is an interferometric technique used to measure the instantaneous velocity of flow fields from changes in phase of a light beam relative to the velocity of a flow-embedded reflecting object [31,32]. Because LDV is non-intrusive, it contends with PIV, which is better suited to define the spatial distribution of the vorticity field [33]. As compared to PIV, time-resolved LDV is more performant in the frequency domain [34].

Here, we propose LDV as a reliable and accurate probe of surface motion to study SWT in viscous fluids. By taking advantage of LDV, we have performed an experimental study considering surface waves excited on different liquid surfaces with discriminated control either by gravity, surface tension or bulk viscosity. From this rationale, we have resolved the SWT dynamics showing that the energy is discretely localized on well-defined lines of the Fourier spectrum. Our results confirm the persistence of this space-time structure in the inertial domain, which is primordial for quantitative analysis of SWT. From the experiments, we perform a test of universality in broad range of constitutive characteristics. Universal scaling has been revealed within SWT cascades created on different fluids at different propagation and amplitude regimes. Finally, the transit from weak to strong turbulence has been also analyzed.

The paper is organized as follows: Section II presents the basic concepts on KZ-scaling theory and SWT-universality necessary to perform the analysis of the experimental results and for setting-up the study in context. Section III describes the experimental rationale. In Section IV, we describe the experimental results obtained for discrete SWT cascades and further analyze them on the framework of the KZ-theory. In Section V, the new results are discussed. An outlook to possible extensions of the work and further applications with an impact in the theory of SWT is given in Section VI. Finally, Section VII summarizes the conclusions.

## II. THEORY

The concept of waving ensemble is central to the theory of wave turbulence [4]. Figure 1A cartoons the idea of discrete ensemble of interacting waves, from which a weak-WT cascade is conceived as the collective behavior of the waving microstates. If enough energy is injected above the critical threshold ($E_{crit}$), the waving levels become progressively interacting through resonant coupling. Consequently, a cascade emerges from the fundamental mode $\omega_0$ up to a border frequency $\omega_{max}$, which is compatible with the "highest occupied" (HO) level. The energy $\epsilon_{HO}$ determines the highest frequency available to the waving ensemble, which defines the bordering cut-off of the cascade. Because energy conservation, the injected energy covers all the possible states in the cascade; consequently $E = \sum_{kl}^{HO} \epsilon_{kl}$. For monochromatic excitation at a forcing frequency $\omega_0$ (and strength $E$), the possible "microstates" correspond to the fundamental mode at $\omega_0$ (with amplitude $a_0$), and the "classical" harmonics (excited at amplitude $a_k \ll a_0$) successively appeared at $\omega_k = k\omega_0$ ($k = 2, 3 \dots$). From a phenomenological point of view, the spectral density of the waving ensemble should appear as a superposition of coupled planar waves as $P(\omega) = \sum_k a_k^2 e^{i\omega_k t} + \sum_{kl} f_{kl} a_k a_l e^{i(\omega_k + \omega_l)t}$ (see Fig. 1B). The waving cascade can either dead rapidly via frictional stress (uncoupled case $f_{kl} = 0$) or undergo amplitude recovery in case of resonant coupling ($f_{kl} \neq 0$) [35]. Cascade dynamics should result from the hybrid energetics of the pure states (with energy $\epsilon_k = a_k^2 \omega_k^2$), combined with the exchange interactions between coupled states ($\epsilon_{kl} = f_{kl} a_k a_l \omega_k \omega_l$, with $f_{kl}$'s being coupling constants). At frequencies $\omega \gg \omega_{max}$, the modes above the "lowest unoccupied" (LU) level cannot be excited anymore (see Fig. 1A); consequently, the cascade ends under frictional stresses so, the ultraviolet catastrophe is observed in the spectral density (see Fig. 1B). No resonances are expected in the uncoupled case ($f_{kl} = 0$), which gives rise to the simplest WT-cascade characterized by an exponential decay of the Kolmogorov's spectrum $P(\omega) \sim e^{-\epsilon} \sim \omega^{-2}$ ($\epsilon \sim a^2 \omega^2$, thus $\alpha = -2$), which leads to frictional death at $\omega_{max} \approx \sqrt{E}$ [35]. However, collective behavior is expected to emerge upon intermodal clustering, which involves second-order coupling interactions between neighbor modes along the cascade; usually triads or tetrads [3,36]. Such a reactive coupling (*exact*- or *quasi*-resonant) might lead to stronger decays in the observed spectrum ($-\alpha \geq 2$) [4], which underlie synergistic distributions of the injected energy up to frequencies higher than expected for uncoupled cascades; these are $\omega_{max} \approx E^\beta$ ($\beta \geq 1/2$) [35-36]. Depending on the specific nature of the coupling interactions, self-organized weak-WT [37], phase locking [38], even synchronization [39], have been observed in this cooperative regime.

In SWT, the KZ-theory establishes the dispersion equation of the surface modes as the key factor determining the pumping mechanism that supports resonant coupling [4,21]. In fluid surfaces, waving motion is expected to be driven by a genuine surface response restored by surface tension ($\sigma$) and gravity ($g$). In terms of the wavevector $k$, the dispersion law is $\omega^2 = (\rho g k + \sigma k^3)/\rho$, which stablishes a crossover from gravity-like dispersion at long wavelengths $\omega_g = \sqrt{gk}$, to capillary ripples at short wavelengths $\omega_c = \sqrt{\sigma k^3/\rho}$ (here, $\rho$ is the fluid density) [40]. The crossover occurs at a capillary frequency $\omega_{cap} = (\rho g^3/\sigma)^{1/4}$, which fixes the boundary between the two propagation regimes. To produce weak SWT, mode coupling occurs as low-order Hamiltonian perturbations between a small number of clustered modes leading to resonance within propagating surface wave packets [41,42]. Specifically, SWT coupling occurs by resonance among nearest neighboring modes [4,41], which establishes the form of KZ-spectra to be dependent on the symmetry of the Hamiltonian but not on the details of energy injection [41,42]. The carrier wave appeared at the forcing frequency $\omega_0$ determines the propagation of the pumped wave packets [41]; only the fact that energy is conserved within these packets is important to evaluate the closure relations that fix mode contributions to the spectral amplitudes [4,21]. Therefore, two different universality classes of discrete SWT are expected depending on the propagation frequency of the forcing mode:

*i) Gravity class g-SWT*: At injection frequencies $\omega_0 \ll \omega_{cap}$, where the leading mode is a long-wavelength gravity wave propagating as $\omega_g \sim k^{1/2}$. In this dispersion domain, the KZ theory assumes a Hamiltonian symmetry that imposes a pumping mechanism dominated by 4-wave interactions; wave-tetrads under the conservation condition $\omega_k + \omega_l = \omega_m + \omega_n$ [4]; the resulting universality class, named *g*-SWT, might be describable as a gravity cascade decaying as $A_h^{(g)} \sim g E_0^{1/3} \omega_h^{-4}$ [21,41]. Consequently, the discrete power spectrum is expected to decay with scaling exponent $\alpha_g = -4$.

*ii) Capillary class c-SWT*. At $\omega_0 \gg \omega_{cap}$, where the leading mode is a capillary ripple dispersing as $\omega_c \sim k^{3/2}$. In this regime, symmetry imposes non-linear coupling to occur as a 3-wave interaction between neighboring triads; these are $\omega_k + \omega_l = \omega_m$ [4]. The resulting scaling exponent is $\alpha_c = -17/6$, which describes this universality class as a capillary cascade decaying as $P_h^{(c)} \sim (\sigma/\rho)^{1/6} E^{1/2} \omega_h^{-17/6}$, and cut-off frequency varying as $\omega_{max}^{(c)} \sim E^{4/3}$ [36,41]. Consequently, the corresponding power spectrum of the discrete modes is expected to decay with scaling exponent $\alpha_c = -17/6$.

### III. EXPERIMENTAL

The laser Doppler velocimeter used in this work (LDV100 Polytec GmbH) focusses a monochromatic laser beam (633nm, 1mW) in a single point of a reflecting surface through of the normal direction. Then, the retro-reflected radiation is collected by the detector to measure the Doppler shift upon surface displacements at variable vertical velocity (see Fig. 1B). This LDV-meter works in time domain by direct measurement of the instantaneous values of the vertical velocities (50 nm/s absolute accuracy at 10 kHz readout). Further spectral analysis is performed in real time by A/D conversion of the analogical signal acquired in time domain followed by Fourier transformation (FFT; VibSoft 5.2 – licensed to Polytec GmbH). The LDV-power spectrum is naturally given for the surface velocities, but the PSD of the vertical displacements can be also obtained by numerical integration eligible from the software toolbox. The fluid is poured in a rectangular container (Teflon; 45 x 150 mm$^2$, 15 mm depth; see Fig. 1C), which is placed on top of an antivibration table (Newport). To reduce spurious vibrations, the velocimeter is mounted in a vertical optical plate installed solidary within the supporting table.

Surface motion is induced as transverse planar waves created under monochromatic excitation with a rectangular blade (stainless steel; 20 mm width, 0.2 mm thick). This mechanical exciter is attached to a capacitive vibrator (Brüel & Kjael 4809), which is powered by a function generator producing highly monochromatic sinusoidal signals (Agilent 33210A). To avoid for high-frequency electronic noising, the output signal is filtered through a band-pass filter (Stanford SR560). The spectral profile of the excitation device was tested using the same LDV probe by reflecting the laser beam in a mirror placed on the bottom of the container. Sharp monochromatic spectral profiles are detected for the pumping channel in the working range of excitation (1-500 Hz). Residual harmonics due to mechanical resonances were detected at specific frequencies, which are systematically discarded. For this set-up, the induced vertical displacement varies linearly with the powering voltage (see Fig. 1D). The laser energy localized in the surface is assumed negligible to cause additional surface motion, a fact experimentally verified since no substantial differences are detectable in the spectra under progressive defocusing and/or lessening the laser beam on the surface.

In this work, we study viscous liquids whose surface waving spans from gravity-controlled to capillary-like response. The propagation distance is estimated by the capillary length $l_{cap} = (\sigma/\rho g)^{1/2}$, which determines the stress balance between gravity and surface tension. All the fluids considered have a similar capillary length ($l_{cap} \approx 2mm$), much smaller than cuvette dimensions, so the surface field is expected to be congruent with a deep-water propagation schema. In time domain, the corresponding capillary frequency is defined as $\omega_{cap} = \sqrt{2g/l_{cap}}$. At excitation frequencies below the capillary frequency ($\omega_0 < \omega_{cap}$), the surface response behaves as gravity-waves. Conversely, at excitation above the capillary frequency ($\omega_0 > \omega_{cap}$), the surface excites capillary ripples. At very low frequencies ($\omega \ll \omega_{cap}$), a collusion of the surface field with the cuvette bottom could occur at large penetration depths; in this gravity regime, a shallow-water description should be required. To test the possible effect of viscous stress on SWT cascades, we studied silicon oils (Paragon Scientific Ltd.) and glycerol solutions (Sigma) with viscosities increasing by three orders of magnitude with respect to water (Milli-Q) (see Table I).

**Table I.** Values of the material properties of the liquids used in this study: density ($\rho$), dynamic viscosity ($\eta$) and surface tension ($\sigma$) ($T = 22^oC$); $l_{cap}$ is the capillary length and $\omega_{cap}$ the capillary frequency that discriminates the regime of surface wave propagation.

| liquid | $\rho$ (g/cm$^3$) | $\eta$ (mPa s) | $\sigma$ (mN/m) | $l_{cap}$ (mm) | $\omega_{cap}$ (Hz) |
|---|---|---|---|---|---|
| water | 1.0 | 0.82 | 72 | 2.7 | 13.6 |
| Hg | 13.6 | 1.6 | 487 | 1.9 | 16.2 |
| oil1 (RTM8) | 0.8 | 10 | 32 | 2.0 | 15.8 |
| oil 2 (RTM14) | 0.8 | 100 | 32 | 2.0 | 15.8 |
| glycerol | 1.3 | 1410 | 63 | 2.2 | 18.8 |

## IV. RESULTS

*Energy cascades: Inertial domain and terminal friction dissipation.* To validate our SWT setup, we worked with surface waves excited on the free surface of viscous liquids. Surface waving was driven at a single frequency $\omega_0$ and excitation amplitude $V_0$, which implies to inject an energy $E_0 = V_0 \omega_0^2$. At low injected energy, below a threshold $E_0 < E_{crit}$, only the linear response is detected as the fundamental mode at $\omega_0$ (see Fig. S1, in *Supplementary Information*). Above this threshold ($E_0 > E_{crit}$), a non-linear cascade of surface modes develops in all the liquids studied. Figure 2 shows the characteristic SWT cascades observed as a succession of discrete harmonics $\omega_k = k\omega_0$, which appear at multiple integers of the fundamental mode excited at $\omega_0$. The spectra are observed to decay with a well-defined discrete-cascade structure $P(\omega) = \sum_h A_h e^{i\omega_h t}$ (see Figs. 2A-C). They contain not only fundamental mode corresponding to the linear response ($h = 1$) but also the subsequent non-linear harmonics ($h \geq 2$). All the excited modes appear as sharp spectral lines (see insets in Figs. 2A-C), as expected for the inertial interval. In all cases studied, we invariably detect SWT as a cascade of discrete harmonics comprised between the excitation frequency $\omega_0$ and the cascade cut-off at $\omega_{max}$. At $\omega > \omega_{max}$, a Lorentzian-like terminal decay is observed as $P(\omega) \sim \omega^{-2}$ (see Fig. 1A), compatible with Brownian noise corresponding to the stochastic (incoherent) surface motions [3]. We observe discrete inertial cascades on the surface of all the liquids studied, independently of their dissimilar properties (like water and Hg shown in Fig. 2); even for highly viscous liquids for which the inertial cascade persists on top of a progressively lower $\omega^{-2}$-background (see Fig. S2 of *Supplementary Information*). As the cascade falls down its potential energy is progressively lost, just till the border frequency where it terminates under a frictional death. All the harmonics are present in the observed cascades (see Fig. 2D), not detecting neither systematic absences nor significant odd-even alternation. Further, they systematically appear as sharp lines insensitive to the increase of frictional stresses, even at increasing frequency (see insets in Fig. 2A-C). The peak widths are found essentially independent of the viscosity of the fluid, as varied by several orders of magnitude (see Fig. S2 of *Supplementary Information*). Further, viscosity does not affect anyway the internal structure of the cascade (see Fig. S3 of *Supplementary Information*). Under continuous excitation, the cascade remains completely stationary; the response ceasing immediately afterward the forcing device is shut-down but being instantaneously recovered upon further excitation. Therefore, cascade energetics is a tradeoff between the pure elastic response of a vibrating membrane under an external force that largely overcomes friction.

*Gravity vs. capillary non-linear cascades: Scale invariance and universality classes.* We are now interested in discriminating the scaling behavior of the detected SWT cascades in dependence to the nature of their pumping mode. SWT cascades were excited at frequencies lying two different propagation regimes separated by the capillary frequency (see Table I). The power of each discrete surface mode $P_h \equiv A_h^{(g,c)}$ is extracted from the experimental amplitudes (see Fig. 2). The results are plotted in Figure 3A, which points out scaling behavior. In order to discriminate each one of the universality classes, the dynamic variables have been conveniently reduced; the power of the harmonics has been normalized by the amplitude of the pumping mode, *i.e.* $A_h^{(g,c)}/A_{h=1}^{(g,c)}$, and the propagation frequencies with respect to the capillary frequency as $\omega_h^{(g,c)}/\omega_{cap}$. Next, we describe separately each class of universality.

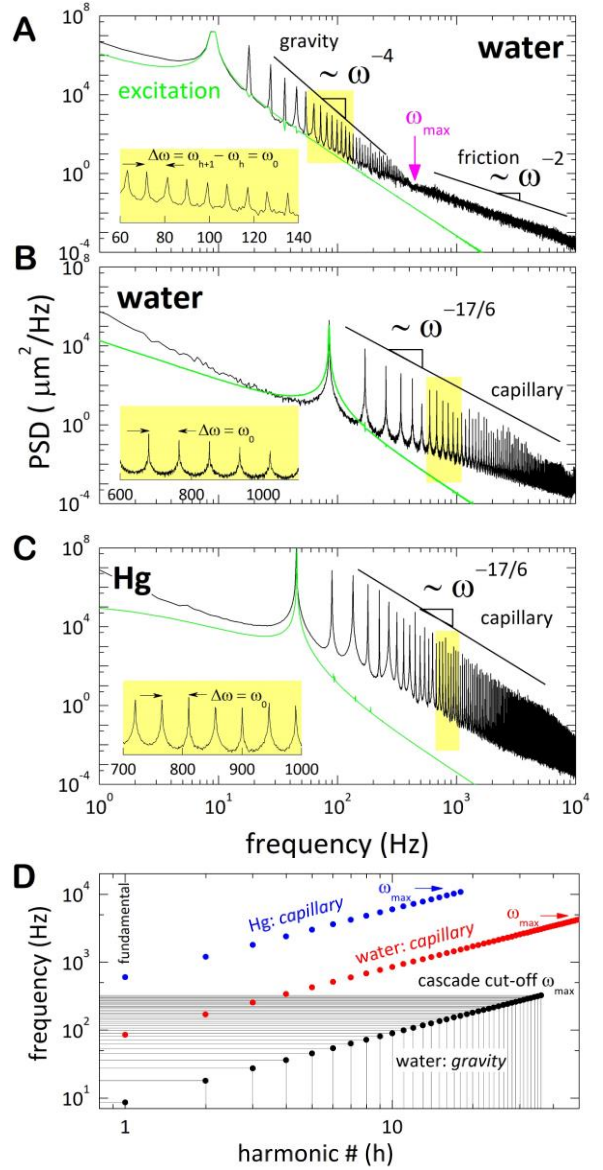

**FIG. 2. Discrete SWT spectral densities:** Typical spectral profiles of gravity-capillary surface wave turbulence excited upon monochromatic pumping at a frequency $\omega_0$ at the free surface of water (panels A-B) and mercury (panel C): Excitation channel (green lines); response channel (black lines), as detected both with the LDV setup. **A)** $g$-SWT spectrum obtained from the surface of water with the $\omega^{-4}$-decay characteristic of gravity-like SWT cascades (at injection frequency $\omega_0 = 8Hz \ll \omega_{cap}$). Terminal Brownian-like noise appears above the cascade cut-off $\omega_{max}$; here $P_{Brown} \sim \omega^{-2}$. **B)** $c$-SWT spectrum obtained from water with the $\omega^{-17/6}$-decay characteristic of capillary-like SWT cascades ($\omega_0 = 90Hz \gg \omega_{cap}$). **C)** $c$-SWT spectrum from mercury ($\omega_0 = 45Hz \gg \omega_{cap}$). The insets zoom the cascade regions dashed in yellow in the main panels; all the harmonics, even the higher, are invariable found with the repetitive discrete structure $\omega_{h+1} - \omega_{h+1} = \omega_0$. **D)** Energy levels corresponding to the discrete cascades of sequential harmonics from the spectral profiles in panels A-C; all the cascades die at a well-defined cut-off $\omega_{max}$, which represents the continuous ceiling that precedes the terminal Brownian tail ($\sim \omega^{-2}$ at $\omega > \omega_{max}$).

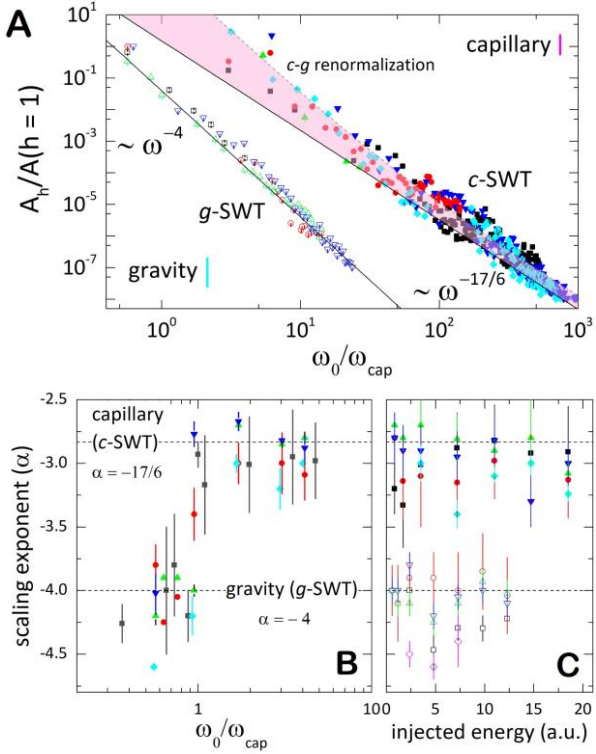

**FIG. 3. Universality classes** of discrete SWT upon monochromatic excitation at $\omega_0$ in different pumping domains: *i)* Gravity-like at $\omega_0 \ll \omega_{cap}$ (*g*-SWT: hollow symbols); and capillary-like at $\omega_0 \gg \omega_{cap}$ (*c*-SWT; solid symbols). **A)** Kolmogorov-like scaling of the spectral amplitudes (vertical bars indicate typical error within each class). Each symbol color represents a different SWT cascade excited on the surface of different liquids (at different frequency and amplitude well within the two pumping domains). The straight lines represent the best fits to a power law $P_h \equiv A_h^{(g,c)} \sim \omega_h^{-\alpha}$ (for all cascades within a same class of universality). The dashed line represents renormalization from *c*-SWT scaling at high frequencies to *g*-SWT scaling at frequencies approaching the capillary boundary from above $\omega_0 \to \omega_{cap}^{(+)}$. Spectral amplitudes within the pink-dashed region are considered somewhat hybrid (*g/c*-SWT). **B)** Renormalization of the scaling exponents from the *g*-SWT class of universality at $\omega_0 \ll \omega_{cap}$ ($\alpha_g = -4$) to the *c*-SWT class at $\omega_0 \gg \omega_{cap}$ ($\alpha_c = -17/6$). **C)** Invariance of the scaling exponents with the injected energy in each regime of wave pumping: gravity (hollow symbols); capillary (solid symbols).

*i) Non-linear gravity waves: Class g-SWT.* At pumping frequencies $\omega_0 \ll \omega_{cap}$, we detect gravity-like cascades with discrete amplitudes nearly decaying as $P_h^{(g)} \sim \omega_h^{-4}$, invariably of the liquid studied (see Fig. 3A; exception of highly viscous fluids for which gravity waves cannot be excited in the experimentally available range of injection energies). From fits to a power-law, the scaling exponent takes the value $\alpha = -4.1 \pm 0.4 \approx \alpha_{grav}$, obtained as an average between all the systems studied. All the observed cascades within this propagation regime belong to the *g*-SWT universality class. The results in Fig. 3A point out how such gravity-like scaling extends out almost two decades above the forcing frequency (see also Fig. 2A). These results suggest the existence of universal scaling invariance within all the *g*-SWT cascades studied. Let's remember, however, that the spectrum becomes continuous at $\omega > \omega_{max}$, where re-scales as $P \sim \omega^{-2}$ (see Fig. 2A). Otherwise said, the gravity cascade terminates into a pure dissipative Brownian tail ($\alpha = -2.1 \pm 0.3 \approx \alpha_{Brown}$), where the inertial domain is no longer sustained. Here, the spectra enter a terminal regime dominated by viscous dissipation; the Brownian dominance extends out with increasing viscosity (see Fig. S4 of *Supplemental Information*). Thus, scaling invariance and universality are both broken beyond the inertial interval where the cascade dies.

*ii) Non-linear capillary waves: Class c-SWT.* From the data in Fig. 3A, we detect capillary-like scaling for the cascades created at pumping frequencies in the capillary domain. At $\omega_0 > \omega_{cap}$, we observe the amplitudes of the discrete modes scaling with a structure $P_h^{(c)} \sim \omega_h^{-17/6}$, as expected for capillary waves (see also Fig. 2B-C). At high excitation frequencies ($\omega_0/\omega_{cap} \gg 1$), the reduced amplitudes are compatible with universal capillary-like scaling, with average exponent $\alpha = -2.9 \pm 0.3 \approx \alpha_{cap}$ observed for different fluids and excitation conditions. This result assigns these cascades to belong to the *c*-SWT universality class. However, systematic deviations are observed at lower excitation frequencies (see Fig. 3); here, the amplitudes depart from capillary-like scaling ($\alpha_{cap} = -17/6$) up to gravity-like behavior ($\alpha_{cap} = -4$) (see Fig. 3A). Such an anomaly suggests class hybridization in approaching the lower boundary of the capillary region ($\omega_0/\omega_{cap} \to 1$). The results in Fig. 3A with *c*-SWT cascades evidence altogether their robustness with respect to *g*-SWT; the capillary waves showing higher relative amplitudes and slower decay than gravity-pumped cascades. Furthermore, inertial *c*-SWT cascades can be excited at high injection frequencies without significant losses (see also Fig. S5 of *Supplemental Information*). The capillary cascades support frictional propagation along larger scales than gravity modes; indeed, the upper borderline of *c*-SWT cascades are blue-shifted with respect to similar *g*-SWT's (Brownian death occurring in the kHz range; see Fig. 2). All these evidences suggest that surface wave packets made of coupled capillary modes are more efficient than gravity waves in propagating frictionless.

*Crossover.* Once the *g*-SWT, or the *c*-SWT cascades have been identified by scaling laws with well-defined exponents, each universality class is recognized to share a common scale invariance. This universal invariance is independent of the constitutive properties of the fluid, which allows us to classify the cascades under a given universality class determined by the propagation characteristics of the pumping mode. Scaling within a given class differ clearly from the other only at well-definite scales far beyond $\omega_{cap}$; however, they become similar as $\omega_{cap}$ is approached. The specific crossover between the two classes has been experimentally explored by obtaining scaling exponents for different cascades excited at variable frequency. Figure 3B collects the results obtained for several cascades excited in different liquids. A crossover from *g*-SWT universality ($\alpha \approx \alpha_g = -4$) to *c*-SWT universality ($\alpha \approx \alpha_g = -17/6$) is clearly detected at the boundary scale $\omega_0/\omega_{cap} \approx 1$; otherwise said, a renormalization between universality classes occurs in switching from gravity to capillary-like dispersion at $\omega_0 \approx$ $\omega_{cap}$. The cooperative character of this renormalization (occurred stepwise) is revealed to be universal, independently of the viscosity (Fig. 3B), or of the amount of energy injected (Fig. 3C).

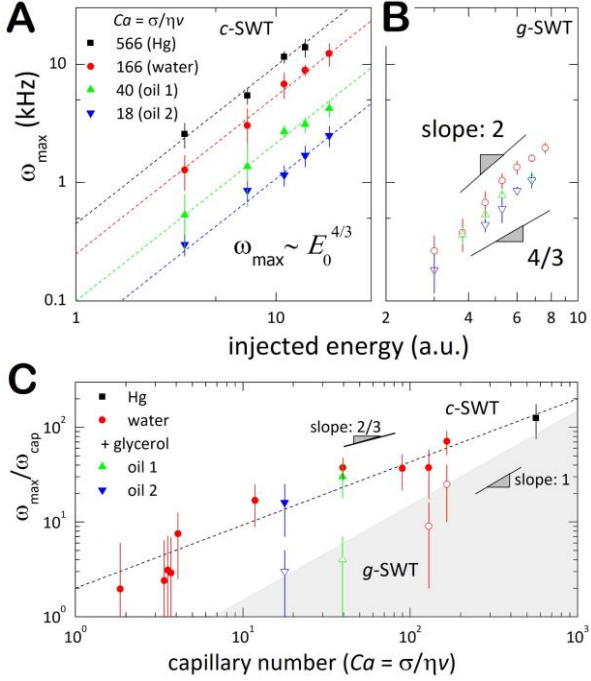

**FIG. 4. Universal size of the inertial domain as determined by the capillary number.** The two panels on top show the scaling of the cut-off frequency on the amount of energy injected to create: **A) Capillary-like** *c*-SWT cascades exhibiting universal Kolmogorov-scaling with the injected energy $\omega_{max}^{(c)} \sim E_0^{4/3}$. Here, the size of the inertial domain decreases as the capillary stress weakens (as determined by the Capillary number $Ca = \sigma/v\eta$; see main text for details). **B) Gravity-like** *g*-SWT cascades with synergistic inertial interval $\omega_{max}^{(g)} \sim E_0^\beta$ with $\beta \geq 4/3$ (see main text). Notice the higher upper boundaries involved in faster capillary cascades than in gravity ones. Because the liquid densities considered are almost equal, the analysis of universality is not possible in this regime (e.g. due to the high density of Hg, we are not able to excite gravity SWT cascades in this liquid). **C) Universality relationship** for the size of the *c*-SWT inertial interval of different viscous liquids (see legend); solid symbols correspond to *c*-SWT cascades ($\omega_0 = 27 Hz > \omega_{cap}$) and hollow symbols to *g*-SWT ($\omega_0 = 10 Hz < \omega_{cap}$). The deformation amplitude is 0.5mm in all cases. The experimental data cover a broad span of viscous-to-capillary stress ratio (as determined by the capillary number $Ca = \sigma/v\eta$). The dashed line represents the universal exponential law found for the *c*-SWT class on the assumption of an energy density completely dominated by capillarity ($\epsilon_c \sim Ca$, so $\omega_{max}^{(c)} \sim E_0^{4/3}$; see main text for details). The grey-dashed region represents the estimated size of the inertial interval determined for gravity cascades on the assumption of $\epsilon_g \sim \omega_{max}^{(g)}$ (under the phenomenological scaling detected for gravity cascades, this is $\omega_{max}^{(g)} \sim E_0^2$; see data in panel B).

*Frictional stress: Cascade death.* Despite the inertial character of the observed cascades, their frictional death has been also clearly pointed out (see Fig. 2, and Fig. S4 of *Supplemental Information*). Such a "ultraviolet catastrophe" is predicted by the KZ theory as a consequence of energy conservation within the waving microstates [3,4,21]. To preserve the energy density, the SWT ensemble should be finite, which necessarily imposes cascade truncation above the HO-level (see Fig. 1A); hence, the cascades fall into a noise background where they cool down by Brownian death ($\sim \omega^{-2}$ at $\omega > \omega_{max}$; see Fig. 2A and Fig. S4 of *Supplemental Information*). The cascades would become heater and broader with increasing the energy injected ($\omega_{max} \sim E^\beta$ with $\beta \geq 1/2$). However, since frictional stresses elicit kinetic slowing down, their upper border is expected to decrease with the bulk viscosity, *i.e.* $\omega_{max} \sim \eta^{-1}$ [1]. The existence of such a viscosity-dependent upper border is concomitant with the ensemble finiteness assumed in the KZ theory, which predicts *quasi*-universal scaling for the border frequency with the injected energy as $\omega_{max} \sim \eta^{-1} E^{4/3}$ [27].

Figure 4 plots the values of the border frequency $\omega_{max}$ detected in experiments with different fluids. The respective data for the two classes of universality show inertial intervals enlarging with the injected energy (see both *c*-SWT in Fig. 4A, and *g*-SWT in Fig. 4B). In the case of capillary cascades, $\omega_{max}$ increases with capillarity (as determined by the capillary number $Ca = \sigma/\eta v$, where $v$ is the intrinsic vertical velocity of the fluid; see caption of Fig. 4 for details). As a matter of fact, the experimental data for *c*-SWT cascades reveal KZ-scaling with the injected energy as $\omega_{max}^{(c)} \sim f(Ca) E_0^\beta$; the average scaling exponent obtained from the best fits to all the experimental data is $\beta = 1.34 \pm 0.12 (\approx \beta_c = 4/3)$, in agreement with KZ-theory. The universality relationship $f(Ca)$ is revealed by Fig. 4C, where all the liquids are observed to collapse into a master function of the capillary number ($Ca$), as expected for the progressive influence of capillary forces in sustaining the inertial domain over viscous friction [4,21]. The linear dependence observed in the double-log plot reveals a phenomenological law $f(Ca) \sim Ca^{2/3}$ for *c*-SWT's, in agreement with a waving energy density completely determined by capillarity, this is $\epsilon_c \sim Ca$; since the coupling interaction of the nonlinear capillary waves is second-order on the exciting field ($\epsilon_c \sim E_0^2$), one expects $\omega_{max}^{(c)} \sim \epsilon_c^{2/3} \sim E_0^{4/3}$, as observed for *c*-SWT cascades (Fig 4A). All in agreement with KH theory [4]. Even more synergistic are the dependencies detected for *g*-SWT cascades. Despite no clear dependence on viscosity is seen for the upper boundary of gravity cascades (see Fig. 4B), the results point to a stronger coupling mechanism than in capillary cascades; for instance, if the energy density for gravity waves involves third-order interactions ($\epsilon_g \sim E_0^3$), one expects $\omega_{max}^{(g)} \sim E_0^\beta$ with $\beta \geq \beta_c$, as suggested by experimental evidence (Fig. 4B and 4C). However, possible troubles with the long-wavelength gravity waves involved prevent for a deeper discussion. From a prudent standpoint, the shorter gravity cascades observed seem compatible with the expected inverse dependence on the viscosity $\omega_{max}^{(g)} \sim \eta^{-1}$ (see Fig. 4C). More advanced work involving larger vessels and higher energy needs to be performed to go further with the quantitative properties of *g*-SWT cascades. Regarding the capillary cascades, however, universal scaling appears clearly pointed out from the current experiments.

*High amplitude: Crossover from weak to strong turbulence.* The limited nonlinear coupling involved in weak turbulence can be eventually overcome at higher energy. Above the critical threshold $E \gg E_{crit}$, a strongly turbulent regime can be eventually entered by involving either, higher unoccupied levels (the LU level and beyond; see Fig. 1A), or higher order interactions [3,4]. The corresponding regime of strong-SWT should appear progressively with increasing $E$ (or increasing $\omega_0$), resulting in a non-universal turbulent surface flow characterized by a high vorticity, the emergence of coherent structures and, eventually, chaotic flow [4,21]. To study scaling along this crossover, we performed experiments with different fluids at varying the energy injected in the surface. The results obtained with $c$-SWT cascades are shown in Fig. 5 (and in Fig. S6 of *Supplementary Information*). Figure 5A shows the spectral profiles of $c$-SWT cascades excited on the free surface of water at increasing injection energy ($E$; similar surface heating is achieved by increasing the forcing frequency at fixed amplitude). Equivalent results were obtained with the other fluids considered (see data for Hg in Fig. S6). The energy injected is understood as "heat" transferred to excite the resonant radiation leading to strong turbulence. At given forcing frequency $\omega_0$, the higher $E$ the faster the surface velocities within the excited cascade, which can be consequently tuned by varying $E(\sim v^2)$. Such an increase of velocity should result in decreasing the ratio between capillarity and frictional stress (as determined by the capillary number, $Ca = \sigma/\eta v$). This frictional control of capillarity is somewhat equivalent, but not exactly equal, to evaluate the degree of turbulence as an increase in the Reynolds number, $Re = \rho vL/\eta$ (see caption in Fig. 5 for details). Because the transit to strong turbulence should occur upon dominance of inertia against capillarity, more adequate control should be performed with the Weber number $We = Re/Ca$, which describes the ratio between the surface densities of kinetic energy and capillary energy per wavelength, this is $We = \rho v^2 L/\sigma$. As a matter of fact, the experimental results in Fig. 5A point out a renormalization from the weak cascades (high $Ca$, low $Re$ and $We \ll 1$), which set-up as "cool" cascades with a broad inertial domain, up to highly-turbulent cascades with a faster decay at high kinetic energy (low $Ca$, high $Re$ and $We \gg 1$), *i.e.* "heater" cascades characterized by a less-capillary inertial domain. When surface motion is tracked, the higher $We$ the stronger the turbulence visually perceptible, first as a disordered surface flow ($We \gtrsim 1$), then followed by a high vorticity at high energies ($We \gg 1$) (similar results are achieved at high amplitudes by increasing the forcing frequency; see movies supplied as *Supplementary Information*).

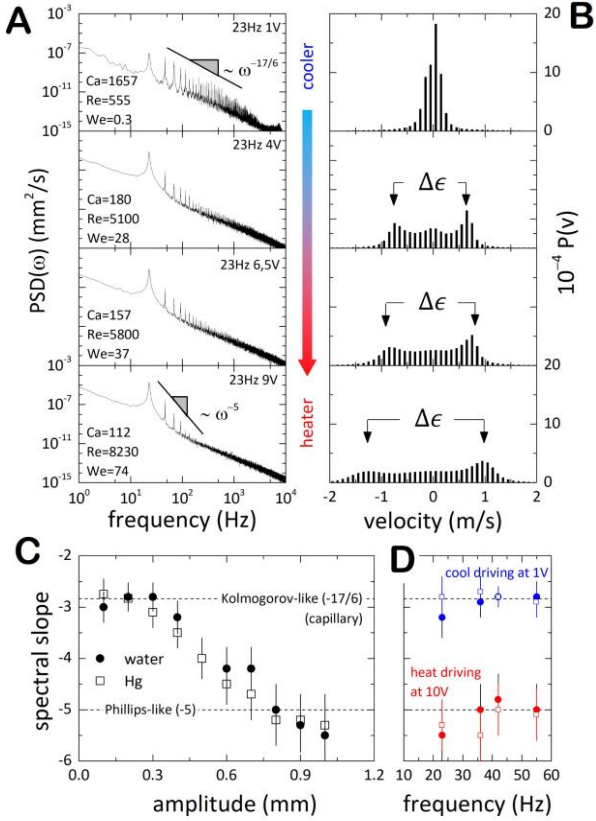

**FIG. 5. Transition from weak $c$-SWT to strong SWT** with increasing the injected energy. The plots represent the case of water; similar results are obtained with other liquids (data not shown; for Hg *e.g.* see Fig. S6 in *SM*). **A) Scaling renormalization** from weak- (top) to strong-turbulence (bottom) with capillarity decreasing at the expense of increasing the frictional stress upon "heater" surface forcing. Because constant surface tension and constant dynamic viscosity are considered, faster wave velocity implies smaller capillarity ($Ca = \sigma/\eta v$), and bigger surface turbulence (as estimated by the Reynolds number ($Re = \rho vL/\eta$), where $L$ is the characteristic length of the surface waving flow (calculated as the wavelength from the dispersion equation). Within these renormalizing spectra, the ratio of inertia *vs.* capillarity is indicated by the Weber number ($We = \rho v^2 L/\sigma = Re/Ca$). The values of the SWT-driving conditions and the resulting dimensionless numbers are indicated in the leftmost panels. **B) Velocity distribution function** as obtained from the histograms of the vertical velocities time-traces corresponding to the spectra in panels A. The probability distributions evolve from Gaussian-like for the cooler cascades (top) to *sin*-like with two lateral maxima as the cascades become heater (bottom). Symmetry breaking with respect to Gaussianity is evaluated as the change $\Delta\epsilon$ in kinetic energy upon heating. **C) Renormalization of the scaling exponents** in the transit from weak- to strong-SWT in water (from the spectra in panels A) and in mercury (from Fig. S6). **D) Scale invariance** at variable frequency for given turbulence status: weak in cool cascades (top); strong in heat cascades (bottom). Data for water appear as solid symbols, and for Hg as hollow symbols.

Figure 5B shows that the observed flow disordering corresponds to progressive departures distribution of the surface velocities from Gaussian to a *sin*-like distribution. A normal Gaussian distribution centered at zero velocity is obtained for "cool" cascades created at relatively low injected energy, *i.e.* $P(v) \sim e^{-\epsilon(v)}$ with hydrodynamic landscape $\epsilon(v) \sim v^2$ at $E \gtrsim E_{crit}$ (see Fig. 5B; upper panel). Upon heating the turbulent cascade, non-linear kinetic energy appears as a broken symmetry around two lateral maxima split $\Delta\epsilon$ towards local basins in the hydrodynamic landscape, *i.e.* $\epsilon(v) \sim v^2 + \Delta\epsilon[o(v^4)]$ at $E \gg E_{crit}$ (see Fig. 5B; with $\Delta\epsilon$ increasing from top to bottom). Because $\Delta\epsilon$ increases upon heating, the data in Fig. 5B suggest velocity redistributions contributing to concentrate energy within a few modes clustered around the master pumping wave, as experimentally observed. Figure 5C shows the experimental values of the scaling exponents calculated from the fits of the cascades in Fig. 5A to a power law. A progressive renormalization is observed from the regular class $c$-SWT ($\alpha_{cap} = -17/6$) to a Phillips-like spectrum ($\alpha_{Phillips} = -5$) compatible with strong turbulence [43,44]. Similar renormalization is observed for water and Hg in terms of the forcing amplitude. In addition, the scaling exponents are found independent of the pumping frequency (at constant injected energy; see Fig. 5D). Despite of the similar behavior depicted by both liquids an ampler study covering a broad range of conditions should be performed for adequate testing of universality. Arguably, if universality holds, a new class of non-trivial mode coupling (other than low-order coupling between nearest neighbors) could be mediating the observed transit from weak to strong turbulence. Given the complexity of this phenomenon, its quantitative analysis deserves detailed attention in a separated piece of work.

## V. DISCUSSION

To the best of our knowledge, prior to this study there was no experimental evidence sufficiently comprehensive to unequivocally state about the universality predicted for weak SWT cascades. Most experimental studies available covered only partial aspects, mainly due to narrow working ranges for energy injection and forcing frequency and very particular choices of material conditions only with certain liquids. Furthermore, the available methods of optical reflectivity [26], or the capacitive probes used in other works [28,45], are limited in terms of absolute precision of vertical velocities and frequency discrimination. These limitations have seriously troubled the detection of discrete SWT cascades, which appear quite as a continuous profile representing a coarse-grained envelope that hides its main dynamical features.

In this work, using LDV for the first time in the SWT context, we have contributed an ample experimental study on the quantitative features of discrete SWT cascades generated upon monochromatic excitation in different viscous liquids under control of gravity, capillary forces and frictional stress. Surface waving has been excited as the inertial response of a system of coupled oscillators pumped at given forcing frequency ($\omega_0$) and variable energy injection ($E$). The presence of SWT as discrete cascades of nonlinear harmonics has been clearly pointed out. Depending on the propagation characteristics of the pumping mode, the observed SWT cascades have been classified within two well-defined classes of universality, which are completely described by well-defined scaling laws [4,21]. The results confirm the picture of weak-SWT as due to the lateral transport of wave packets created by a carrier mode (the fundamental mode of frequency $\omega_0$ and all its harmonics). All these nonlinear modes formally constitute a waving ensemble sustained by the stationary driving force. Consequently, a waving packet of resonant radiation is formed at the surface, which retains the propagation properties of the pumping mode as the universality class that define the Hamiltonian interaction within the ensemble. The results essentially agree with the KZ theory of weak-SWT [4,21], which was formally developed as a particular 2D-Kolmogorov's theory of weak turbulence [3].

Regarding gravity waves, the only support of KZ's theory was for many years the early experiments on sea-waves reported by Toba and Mitsuyasu [46]. Using forcing wind, they excited very long water waves in a relatively narrow channel. The natural set-up considered lacked spatial isotropy and imposed a polychromatic excitation, two upsetting factors inflicting trouble for adequate analysis of $g$-SWT features. Getting further insight on the turbulence associated to water waves is essential to understand non-linear sea dynamics [4], particularly in prediction analysis of catastrophic oceanic events [47]. Because the practical relevance of sea waving motion in ocean physics, the study of universal gravity cascades particularly those leading to rogue waves has recovered both applied and theoretical interest [47]. Most of the experimental studies available to date for $g$-SWT deal with random pumping able to set-up wind-like excitation (see Ref. [21] for a review). The only experimental evidence on the excitation of gravity cascades has been reported by Falcon et al. [28]. However, that study was also performed under polychromatic excitation giving rise to continuous cascades where the specific nature of the involved modes cannot be specified. The continuous cascades recreated in that work show indeed signs of hybridization, because excited in the gravity domain they renormalize to capillary-like at a crossover occurred above the capillary frequency. Such hybridization could arise from bound waves hypothesized by Zakharov to share the hydrodynamic field between the excited modes and non-resonant slave modes appeared in the continuous cascade [48]. The existence of KZ-gravity turbulence was later demonstrated by Cobelli et al. [49], again in experiments performed under broadband excitation. Although the existence of an inertial regime that is dependent on the forcing amplitude was evidenced in that work [49], they were unable to determine the control parameters relevant to describe the weak regime of gravity-like turbulence. Therefore, no unequivocal proof of $g$-SWT universality has been still pointed out from the evidence available [28,49]. Our results, however performed under monochromatic surface excitation and with LVD detection, clearly point out the existence of genuine $g$-SWT cascades that, being pumped by a leading mode that propagates gravity-like, can excite higher harmonics even entering the capillary domain. Interestingly, our observations of the $\omega^{-4}$-law in discrete $g$-SWT cascades represent a first demonstration of the existence of a universal hierarchy of coupling interactions inside packets of non-linear gravity waves. Nevertheless, only partial agreement with KZ theory has been achieved in this preliminary study, probably due to resolution limitations. Further work with the LDV device could enlighten the nature of the coupling mode interactions within the $g$-SWT class of universality (hypothesized to deal with wave tetrads) [4].

Comparatively to gravity cascades, the $c$-SWT capillary class has been amply studied especially with liquid mercury [21,28] and water [21,23,27,50]. The existence of continuous cascades of capillary turbulence exhibiting $\omega^{-17/6}$-scaling was demonstrated by Falcon et al. [28] in experiments under broadband excitation. Almost all published studies deal with continuous cascades under random or broadband excitation, with the exception of Brazhnikov et al. [27], who reported the observation of discrete capillary cascades on water. Those authors took advantage of the diffusing-light technique previously developed in a similar study with liquid hydrogen [26]. Despite the power-law decay of the discrete cascades revealed in those experiments, no definite evidence of scaling behavior could be raised since the inertial domain appeared to be spuriously contaminated by parasite signals (probably arising from the complex diffraction pattern of the grazing light beam used in those experiments). From the evidence accumulated, reasonable agreement with the KZ theory has been pointed out for capillary cascades (see Refs. [4,21] for comprehensive reviews). Among the studies dealing with discrete capillary waves only Brazhnikov et al. [26,27] have reported an inertial domain as a power-law with an exponent of near to $-17/6$, followed by a cut-off frequency that increases with the amplitude of pumping. Using a similar observational method, Punzmann et al. [51] have analyzed capillary mode interactions at the crossover to the continuum, a setting that allowed for studying wave coherence to lead three-wave phase coupling. However, that study was performed via parametric excitation (the driving force acts over the whole container), which makes difficult to interpret the result as corresponding to a genuine surface turbulent cascade. Despite the accumulated evidence, however, the exact nature of mode coupling within the *c*-SWT class remains still a matter of theoretical conjecture. To explore such a coupling localized harmonic pumping with sum-frequency excitation might be chosen in discrete ensembles. In the theoretical side, the energy transported within continuous capillary cascades has been analyzed with a master kinetic equation imposing steady-state, scale-invariant, solutions that describe a bidirectional flux of energy from the injection source towards other scales [52]. However, the fitting range of these scaling predictions is usually restricted in experimental studies, where the discrete harmonics are often non-resolved, which implies large errors in the measured exponents [52]. Particularly undermined by experimental inaccuracy are the stepper coupling that defines the inverse cascades, with scaling exponent $\alpha_{cap}^{(-)} = -11/3$ [53], which is relatively higher than the slower decay of the direct capillary cascade $\alpha_{cap}^{(+)} = -17/6$.

To detect surface waving cascades, either direct or inverse, most experimental literature deals with capacitive probes [28], and with laser deflection methods [26,27], which are somewhat troubled by limitations in resolution. Our study with LDV using an ample collection of viscous liquids demonstrates, by the first time, that discrete *c*-SWT cascades made of capillary waves do exist at fluid interfaces. Our results appear because of resonant coupling between discrete modes pumped by a carrier capillary wave whose propagation characteristics impose the structure of the *c*-SWT cascade, a phenomenon that seems to be universal in terms of the capillary number. The spectral characteristics resolved at high-accuracy by LDV represent a promise for insightful experiments with capillary cascades in fluids of different nature. In general, our approach using LDV to probe SWT in fluid surfaces overcomes much of the experimental difficulties of previous methods [26-28], having provided a reliable and powerful rationale to resolve the fine structure of the discrete cascades constituted by single harmonics under monochromatic excitation.

The examples shown in Figure 2 typify as high-resolution the LDV spectral analysis of the SWT cascades constituted by discrete modes, so we advocate for the powerfulness of our experimental approach. The observed cascades have been quantitatively analyzed in terms of scaling, universality and renormalization within their own classes of universality. Furthermore, we have experimentally demonstrated that these classes are genuinely defined by the propagation characteristics of the pumping mode, which seems to work as the carrier of the non-linear packet of surface waves that generates turbulence. The fine structure of the waving interactions within these packets constitutes a major question that could be addressed in the future with the proposed rationale using LDV. The precise validation of the general theories of WT could become a reality in the experimental setting here proposed: focused excitation of monochromatic surface radiation and LVD detection.

## VI. OUTLOOK

From a fundamental standpoint, studies in SWT are extremely interesting as share the basic nonlinear models of waving interactions, such as the nonlinear Schrödinger and nonlinear Klein-Gordon equations, as well as the field theory approaches, *e.g.* Feynman-type diagrams of classical interactions between waving *quasi*-particles. Other important links exist with the kinetic theory and stochastic partial differential equations. All these connections have refreshed recent interests in SWT systems across the physics and the mathematics communities [21]. The classical Kolmogorov's theory of weak turbulence is based upon the idea that the energy flows frictionless across the waving ensemble through weak-coupling interactions assumed to exist between triads, or tetrads in some cases, of neighboring modes [2-4]. These weak couplings keep the first perturbative terms of the system's Hamiltonian.

Building upon these classical ideas, a field theory has been recently formulated for weak-WT, namely, the MMT theory [54]. Approaching from the non-linear Schrödinger equation, the MMT theory gets dispersive-wave features for which organized turbulence holds [54,55]. Under the assumption of weak-coupling between neighboring modes, resonance appears in MMT to be dependent on the specific nature of the hydrodynamic interactions, which are mainly fixed by the pumping mode [54,55]; consequently, the propagation of the forcing mode should determine the way on how the cascade falls down, as experimentally demonstrated in this work.

The SWT- rationale here proposed, with discrete cascades or another, could become a benchmark to validate these theories in the experimental setting. With the experimental tool now available, the presence of Hamiltonian symmetries can be efficiently explored by high-resolution spectral analysis of the harmonic overtones in the discrete cascades. To explore the coupling mechanisms that pump the cascades within every universality class, we can eventually excite packets of surface radiation potentially able to interact each other under control of driving frequency, driving amplitude and boundary conditions. Furthermore, the transition from weak to strong SWT can be efficiently explored in our experimental setting. Indeed, the breakdown of the "weak" limit has been evidenced to occur beyond a critical ceiling. A symmetry breaking has been observed to occur in entering the strongly turbulent regime; the stronger the turbulence the longer the intermodal coupling that breaks the symmetry of the weakly-perturbed Hamiltonian, which keeps higher-order interactions that result into an appreciable vorticity. This symmetry-breaking has been clearly pointed by our preliminary experiments, which encourage us to explore far beyond this territory. The resulting cascades of strong-SWT are expected indeed to be highly non-universal [44], a question that might be properly studied in experiments. Strong-WT is also predicted to undergo coherent flow, under some particular conditions leading to highly-structured vortices [56], but to hydrodynamic chaos in others [57]. Unfortunately, there is no systematic theory for strong WT, although there may arise states that satisfy a simple scaling principle, the so-called critical balance [58], which states a saturation of the energy spectrum when the nonlinear interaction becomes slower than the driving wave. The MMT theory predicts a continuous transition from weak to strong turbulence [54, 59,60]. All these concepts can, at least potentially, be explored by using discrete SWT cascades probed with the LDV method here proposed.

## VII. CONCLUSIONS

Using laser Doppler velocimetry (LDV), the existence of discrete wave cascades has been pointed out on fluid interfaces upon monochromatic excitation. The use of viscous liquids imposing frictional stresses has allowed for identifying the surface wave cascade as a pure inertial response that terminates into a friction-governed Brownian noise. Two universality classes of surface wave turbulence have been identified to follow well defined scaling laws with the time scale as determined by frequency, $g$-SWT with $\omega^{-4}$-decay for gravity waves and $c$-SWT with $\omega^{-17/6}$-decay for capillary waves. With the surface tension spanning from the high value of mercury down to the low tension of organic liquids, and viscosity varying across more than three decades, the universality features of weak-SWT have been pointed out to be describable in terms of dimensionless numbers, specifically the capillary number for the inertial domain of $c$-SWT and the Weber number for the transit from weak- to strong-SWT. Therefore, LDV has been revealed as a powerful tool for studying SWT.

## ACKNOWLEDGMENTS

This research was partially supported by Ministerio de Investigación, Innovacion y Universidades under grant FIS2015-70339 and by Comunidad de Madrid under grants S2013 / MIT-2807, P2018/NMT4389 and Y2018/BIO5207. We thank our colleague Dr. E. Montoya, who provided expertise with the LDV device, and Prof. J. Fernández-Castillo for generosity in free leasing laboratory space. We are also immensely grateful to Prof. M.G. Velarde (Instituto Pluridisciplinar – Universidad Complutense de Madrid) for critical discussions on the results and discerning comments on the manuscript.

[1] L.D. Landau and E.M. Lifshitz. *Fluid Mechanics* (Pergamon Press, New York, 1959).
[2] A. N. Kolmogorov, Dokl. Akad. Nauk SSSR 30, 299 (1941).
[3] V. E. Zakharov, V. S. L'vov, and G. Falkovich, *Kolmogorov Spectra of Turbulence I* (Springer, Berlin, 1992).
[4] S. Nazarenko, *Wave Turbulence* (Springer-Verlag, Heidelberg, 2011).
[5] A.C. Newell and B Rumpf. Wave turbulence. Rev. Fluid Mech 43, 59 (2011).
[6] G. Falkovich and K.R. Sreenivasan, Physics Today 59, 43 (2006).
[7] V. Zakharov, F. Dias and A. Pushkarev. One-dimensional wave turbulence. Phys. Rep. 398(1), 1 (2004).
[8] T. Peacock and G. Haller. Lagrangian coherent structures: the hidden skeleton of fluid flows. Phys. Today 66(2), 41 (2013).
[9] G. Falkovich, K. Gawȩdzki, and M. Vergassola. Particles and fields in fluid turbulence. Rev. Mod. Phys. 73, 913 (2001).
[10] V.I. Arnold and B.A. Khesin. *Topological Methods in Hydrodynamics* (Springer-Verlag, Heidelberg, 1998).
[11] Gamba, I. M., Smith, L. M., & Tran, M. B., On the wave turbulence theory for stratified flows in the ocean (2017).
[12] Vallis, G. K., *Atmospheric and oceanic fluid dynamics* (Cambridge University Press, 2017).
[13] G. S. Bisnovatyi-Kogan and S. A. Silich, Rev. Mod. Phys. 67, 661 (1995).
[14] D. J. Southwood, Nature (London) 271, 309 (1978).
[15] L'vov, V. S., & Nazarenko, S., Spectrum of Kelvin-wave turbulence in superfluids (2010).
[16] Grach, S. M., Electromagnetic radiation from artificial ionospheric plasma turbulence (1985)
[17] P. Tabeling, Phys. Rep. 362, 1 (2002).
[18] G. Falkovich and K.R. Sreenivasan, Physics Today 59, 43 (2006).
[19] P. Tabeling, Phys. Rep. 362, 1 (2002).
[20] V. Zakharov. Eur. J. Mech. B 18:327–44 (1999).
[21] S. Nazarenko and S. Lukaschuk. Wave Turbulence on Water. Annu. Rev. Condens. Matter Phys. 7, 61-88 (2016).
[22] R.G. Holt and E.H. Trinh. Phys. Rev. Lett. 77, 1274 (1996).
[23] W.B. Wright, R. Budakian, D.J. Pine and S.J. Putterman. Science 278, 1609 (1997).
[24] G.V. Kolmakov. Phys. Rev. Lett. 93, 074501 (2004).
[25] E. Falcon, C. Laroche and S. Fauvé. Phys. Rev. Lett. 094503 (2007).
[26] M.Y. Brazhnikov, G.V. Kolmakov, A.A. Levchenko and L.P. Mezhov-Deglin. Capillary turbulence at the surface of liquid hydrogen. JETP Lett. 73, 398 (2001).
[27] M.Y. Brazhnikov, G.V. Kolmakov, A.A. Levchenko and L.P. Mezhov-Deglin. Observation of capillary turbulence on the water surface in a wide range of frequencies. Europhys. Lett. 58, 510 (2002).
[28] E. Falcon, C. Laroche and S. Fauve. Observation of gravity-capillary wave turbulence. Phys. Rev. Lett. 98:094503 (2007).
[29] G.V. Kolmakov, P.V. Elsmere McClintock and S.V. Nazarenko. Wave turbulence in quantum fluids. Proc. Nat. Acad. Sci. USA 111 (1) 4727 (2014).
[30] R. Lindken and W. Merzkirch. A novel PIV technique for measurements in multiphase flows and its application to two-phase bubbly flows. Experiments in Fluids 33, 814 (2002).
[31] L.E. Drain. The Laser Doppler Technique (John Wiley & Sons, 1980).
[32] Y. Yeh and H.Z. Cummins. Localized Fluid Flow Measurements with a He-Ne Laser Spectrometer. Applied Physics Letters. 4 (10): 176 (1964).
[33] T. Saga, H. Hu, T. Kobayashi, S. Murata, K. Okamoto and S. Nishio. A comparative study of the PIV and LDV measurements on a self-induced sloshing flow. J. Visualiz. 3, 145–156 (2000).
[34] Dalziel, S. B., Carr, M., Sveen, J. K., & Davies, P. A. Simultaneous synthetic schlieren and PIV measurements for internal solitary waves (2007).
[35] K. Kaneko. *Theory and Applications of Coupled Map Lattices* (Wiley, New York, 1993).
[36] V.E. Zakharov and N.N. Filonenko. J. Appl. Mech. Tech. Phys. 8, 37 (1967).
[37] U. Frisch, *Turbulence* (Cambridge University Press, Cambridge, 1995).
[38] S.H. Strogatz and R.E. Mirollo, *Phase-locking and critical phenomena in lattices of coupled nonlinear oscillators with random intrinsic frequencies*, Physica D 31, 143-168 (1988).
[39] A. S. Pikovsky, M. G. Rosenblum, and J. Kurths, *Synchronization: A Universal Concept in Nonlinear Sciences*, Cambridge Non-linear Science Series (Cambridge University Press, London, 2003).
[40] F. Monroy, F. Ortega, R.G. Rubio and M.G. Velarde. Surface rheology, equilibrium and dynamic features at interfaces, with emphasis on efficient tools for probing polymer dynamics at interfaces. Adv. Colloid Interf. Sci. 134, 175 (2007).
[41] F. Dias and C. Kharif. Nonlinear gravity and capillary-gravity waves. Annu. Rev. Fluid Mech. 31, 301 (1999)

[42] E. Kartashova. Exact and quasi-resonances in discrete water wave turbulence. Phys. Rev. Lett. 98, 214502 (2007)

[43] B.B. Kadomstev. *Plasma Turbulence* (Academic, New York, 1965).

[44] P.A. Robinson. Nonlinear wave collapse and strong turbulence. Rev. Mod. Phys. 69(2), 507 (1997).

[45] B. Issenmann and E. Falcon. Gravity wave turbulence revealed by horizontal vibrations of the container. Phys Rev E. 87 1, 011001 (2013).

[46] Y. Toba and H. Mitsuyasu. *Ocean surface waves* (Reidel Publihsing Co., Dordrecht, 1985).

[47] E. Pelinovsky and C. Kharif. *Extreme Ocean Waves* (Springer, London, 2016).

[48] V. E. Zakharov. Phys. Scr. T142, 014052 (2010).

[49] P. Cobelli, A. Przadka, and P. Petitjeans. Different regimes for water wave turbulence. Phys. Rev. Lett. 107, 214503 (2011).

[50] W.B. Wright, R. Budakian and S.J. Putterman. Phys. Rev. Lett. 76, 4528 (1996).

[51] H. Punzmann, M.G. Shats and H. Xia. Phase randomization of three-wave Interactions in capillary waves. Phys. Rev. Lett. 103, 064502 (2009).

[52] L.V. Abdurakhimov, M. Arefin, G.V. Kolmakov, A.A. Levchenko, Y.V. Lvov and I.A. Remizov. Bidirectional energy cascade in surface capillary waves. Phys. Rev. E. 91, 023021 (2015).

[53] L.V. Abdurakhimov, M. Arefin, G.V. Kolmakov, A. Levchenko, Y.V. Lvov and I.A. Remizov. Bidirectional energy cascade in surface capillary waves. Phys. Rev. E 91, 023021 (2015).

[54] A.J. Majda, D.W. McLaughlin and E.G. Tabak. A one-dimensional model for dispersion wave turbulence. J. Nonlinear Sci. 7, 9 (1997).

[55] S. Dyachenko, A.C. Newell, A. Pushkarev and V. E. Zakharov. Optical turbulence: Weak turbulence, condensates and collapsing filaments in the nonlinear Schrödinger equation. Physica D 57, 96 (1992).

[56] J. Sommeria, S.D. Meyers and H.L. Swinney. Nature 337, 58 (1989).

[57] P. Berge, Y. Pomeau and Ch. Vidal, *L'Order dans le Chaos* (Hermanm Paris, 1984).

[58] S.V. Nazarenko, A.A. Schekochihin. J. Fluid Mech. 677:134–53 (2011).

[59] D. Cai, A.J. Majda, D.W. McLaughlin, E.G. Tabak. Dispersive wave turbulence in one dimension. Physica D 152–153, 551 (2001).

[60] S. Chibbaro, F. De Lilloand M. Onorato. Weak versus strong wave turbulence in the Majda-McLaughlin-Tabak model. Phys. Rev. Fluids 2, 052603(R) (2017).

# Universality classes of surface wave turbulence as probed by laser Doppler velocimetry in viscous fluids.

Mikheil Kharbedia[1] and Francisco Monroy[1,2,*]

# Supplementary Information

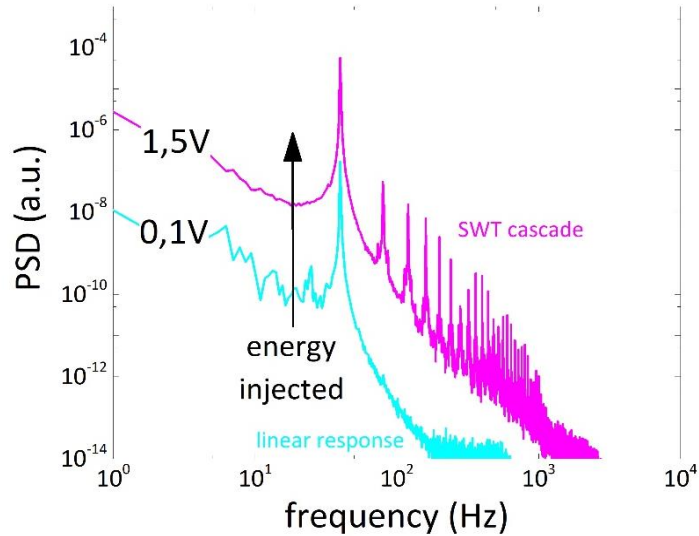

**Figure S1. Spectral profiles of surface waves.** Linear and nonlinear waving response of the water surface in the capillary regime ($c = 40\ Hz$). Only the pure-linear response at the fundamental frequency is observed with injecting low energy (approximately 0.2 mm surface deformation corresponding to a driving voltage of 0.1V; cyan line). A nonlinear SWT cascade composed by progressive harmonics appear upon Increasing the injected energy by a factor near 10 (approximately 0.7 mm amplitude at 1.5V; magenta line).

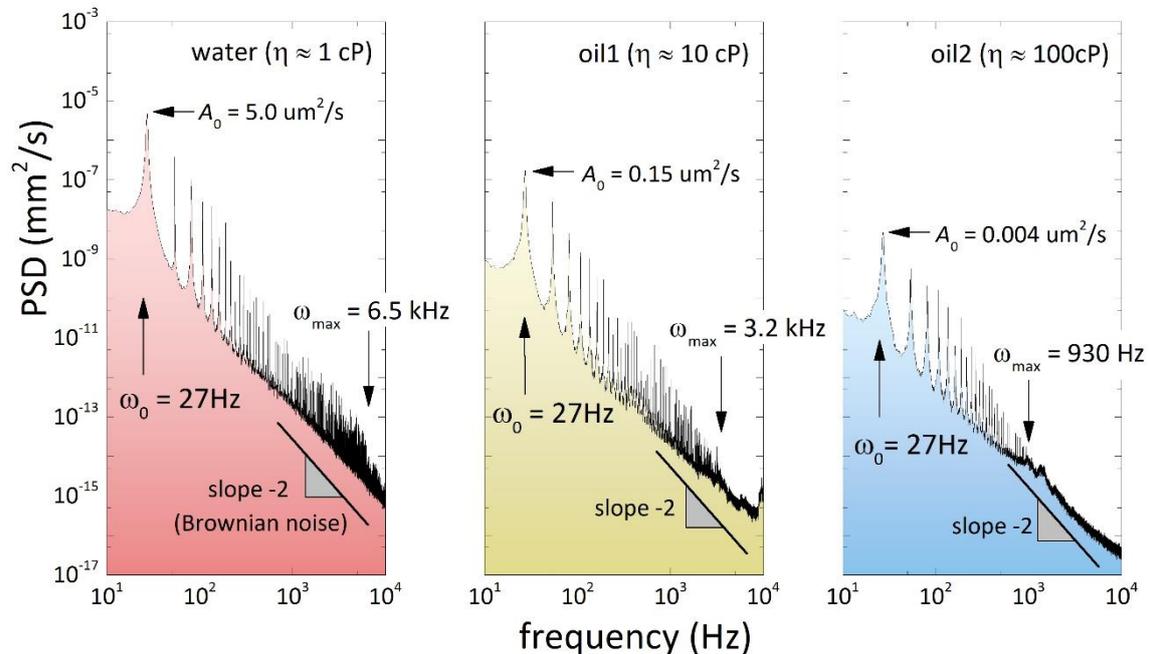

**Figure S2. SWT as an energy tradeoff between elastic response and frictional leakage.** Due to the pure elastic (inertial) response of the surface as a vibrating membrane upon external forcing by energy injection, all the SWT cascades appear as a complete sequence of discrete modes with an inertial character (narrow peaks; non-dissipative), which are mounted on top of a continuous background corresponding to the stochastic (Brownian) motions of the fluid surface. Such a background is obviously

regulated by viscous friction, which imposes the characteristic $\omega^{-2}$-dependence to the dissipated power. Consequently, the higher the viscosity the lower the background (see panels from left (water) to right (high viscous oils)). Also due to friction dominance over inertia, above a critical frequency ($\omega_{max}$) the cascade breaks down into a continuous tail that extends out as the $\omega^{-2}$-noise due to friction. As a proof of the control by friction, the border frequency $\omega_{max}$ is clearly observed to decrease with increasing viscosity (from left to right).

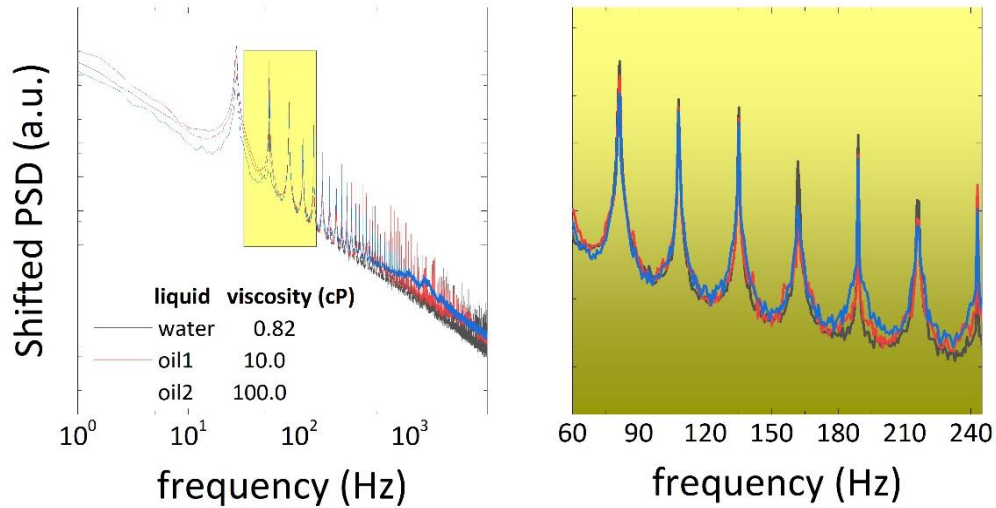

**Figure S3. Inviscid character of the inertial domain.** Universal superposition of the SWT spectral profiles obtained from liquids of very different dynamic viscosity (shifted vertically till amplitude superposition): 1) water with $\eta = 0.82 cP$ (black line); 2) silicon oil 1 with $\eta = 10 cP$ (red line); and 3) silicon oil 2 with $\eta = 100 cP$ (blue line) (excitation frequency in the capillary domain; $\omega_0 = 40\ Hz \gg \omega_{cap}$). The zoom in the rightmost panel points out absolute peak superposition (equal peak areas; no dissipative peak broadening), as expected for a same class of capillary universality (*c*-SWT, following the nomenclature in the main text).

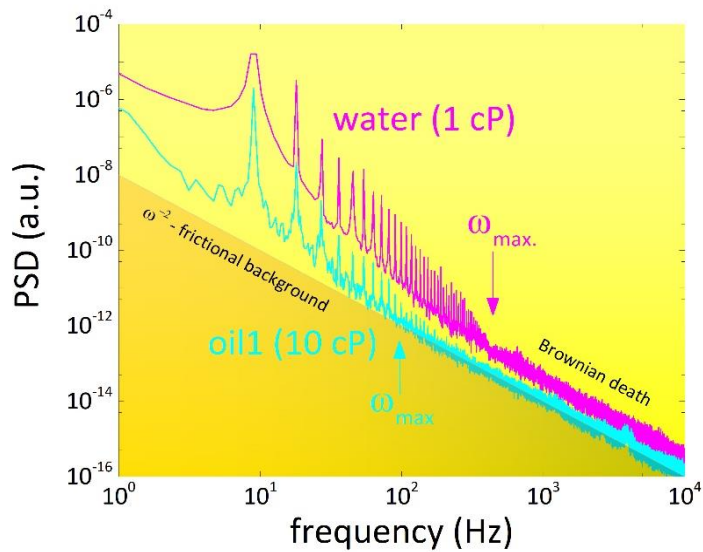

**Figure S4. Frictional (Brownian) death above the border frequency:** Comparison between the SWT spectra for water ($\eta = 0.82 cP$) and a viscous oil ($\eta = 10 cP$) by increasing the amount of frictional stress (by a factor 10), which kills the more viscous

cascade at a frequency 10 times lower than in water. In oil the inertial interval breaks down at lower frequency than in water, due to higher viscosity. Above $\omega_{max}$ the inertial response is completely overcome by friction in an ultraviolet catastrophe; here, the cascade terminates to fall down into the Brownian noise that define the frictional background.

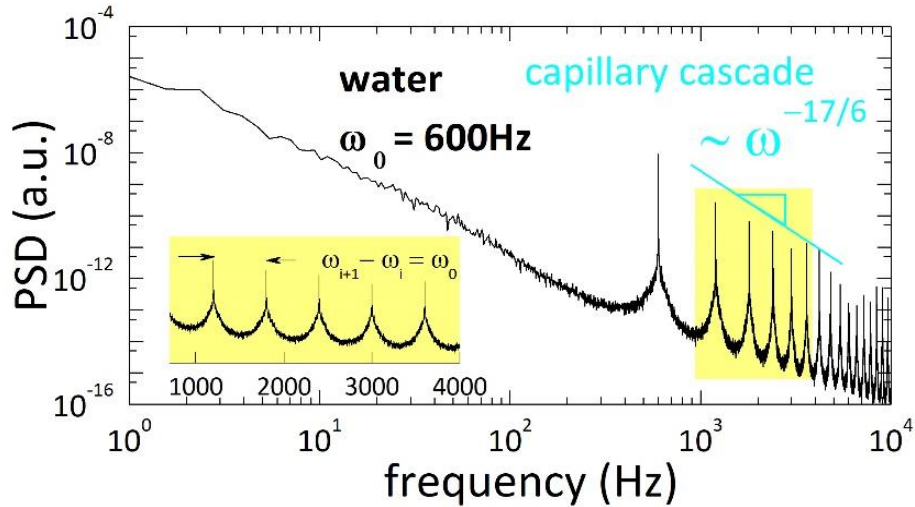

**Figure S5. High-frequency *c*-SWT cascade with a frictionless-preserved inertial domain.** This representative example with water (obtained at excitation frequency $\omega_0 = 600\ Hz \gg \omega_{cap}$) shows the existence of inertial cascades of discrete harmonics pumped from the fundamental mode even at very high frequencies much higher than the capillary frequency. The cascade remains perfectly inviscid, despite the presence of frictional stresses that could eventually dominate this regime in case of pumping at lower frequencies. Despite the different frequency domain for non-linear mode coupling, the mechanism of energy pumping remains unaltered in this scale, so there is no essential difference with the same class of turbulence created at lower frequencies.

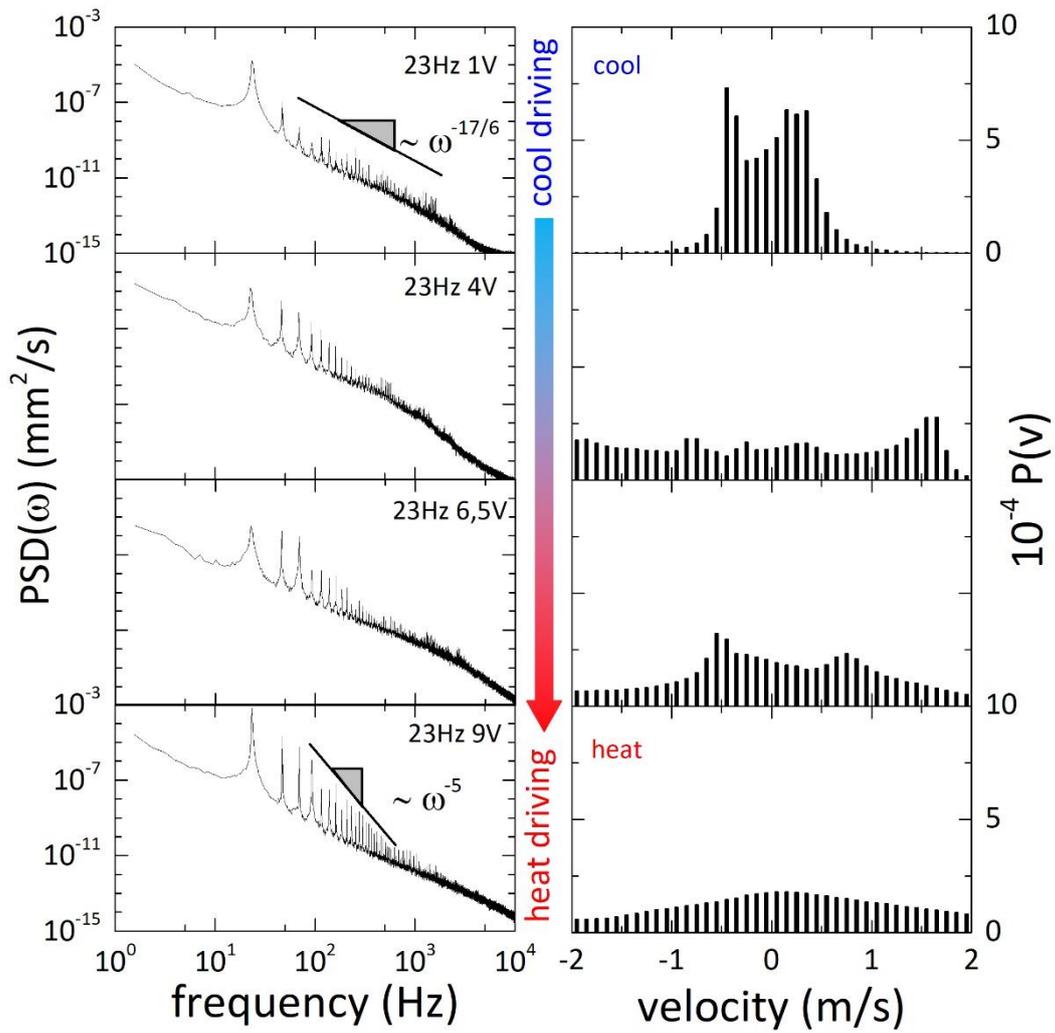

**Figure S6. Transition from weak- to strong-SWT in liquid mercury.** The data corresponds to the same experiment as reported for water in Fig. 5 of the main text (see legend there).